\documentclass[psfig,amsmath,amssymb,12pt]{article}
\usepackage[T1]{fontenc}
\usepackage{graphicx}\usepackage{epsfig}\usepackage{portland}
\usepackage{amsmath}\usepackage{amssymb}\usepackage{amsfonts}
\usepackage{bm}
\usepackage{exscale}\usepackage{rotating}
\usepackage{dcolumn}
\usepackage{color}

\usepackage{fancyhdr}
\fancypagestyle{plain}{%
	\fancyhead{} 
}

\begin{document}
\title{
STRUCTURAL EFFECTS IN THE NUCLIDE DISTRIBUTIONS OF THE       
RESIDUES OF HIGHLY EXCITED SYSTEMS
}
%
%
\date{}
\maketitle
%
%
\vspace{-1.5cm}
\begin{center}
\footnotesize P. NAPOLITANI, F. REJMUND and L. TASSAN-GOT
\\\vspace{1ex}
IPN Orsay, IN2P3, 91406 Orsay, France
\bigskip\\
\footnotesize M. V. RICCIARDI, A. KELIC, K.-H. SCHMIDT and O. YORDANOV
\\\vspace{1ex}
GSI, Planckstr. 1, 64291 Darmstadt, Germany
\bigskip\\
\footnotesize A. V. IGNATYUK
\\\vspace{1ex}
IPPE, Bondarenko Squ. 1, 249020 Obninsk, Russia
\bigskip\\
\footnotesize C. VILLAGRASA
\\\vspace{1ex}
DAPNIA/SPhN, CEA/Saclay, 91191 Gif sur Yvette, France
\end{center}

%
%
\vspace{1.cm}
\begin{tabular}{|p{10.cm}|}
New data from GSI on the production-cross-section for fragmentation
of the systems $^{56}$Fe+$p$ and $^{56}$Fe+$^{\textrm{nat}}$Ti at 1~$A$~GeV
revealed the appearance of even-odd staggering in the cross-section
distribution for chains of isotopes with given $N-Z$.
The staggering is strongly enhanced for the chain $N=Z$,
it reduces as the production moves away from the $N=Z$ chain,
and it reverses for the most neutron-rich odd-$A$ residues.
These phenomena, observed in the residues of rather violent
reactions, are related to structural effects in the level-densities
below the particle-emission threshold.
\end{tabular}
\vspace{1.cm}

\section{Introduction}

\begin{figure}[t!]
\begin{center}
\includegraphics[angle=0, width=1\textwidth]{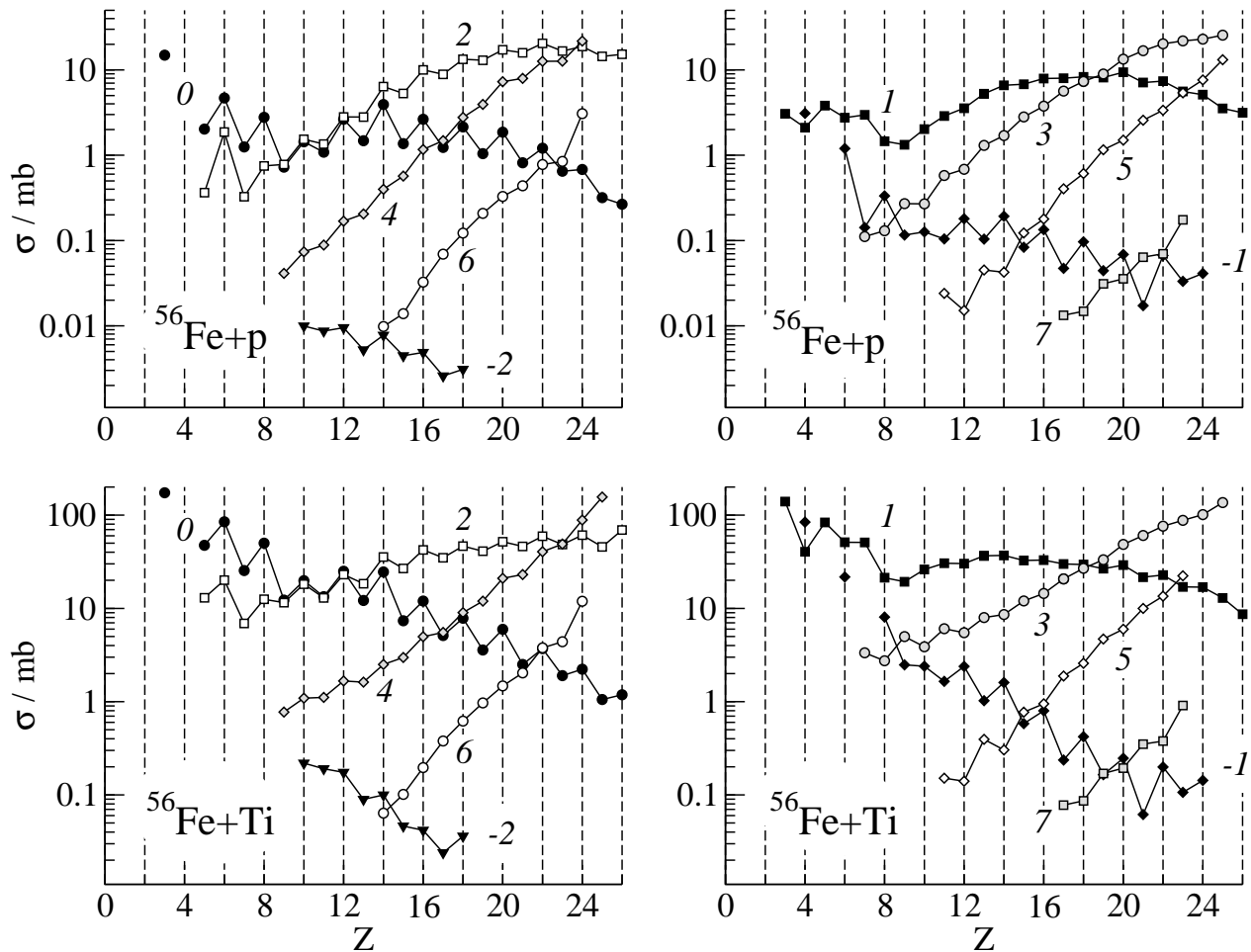}
\end{center}
\caption
{
Experimantal cross sections of $^{56}$Fe+$p$ (top)
and $^{56}$Fe+$^{\textrm{nat}}$Ti (bottom) for even-mass residues
(left) and odd-mass residues (right), respectively.                         
The cross sections are ordered in chains according to given
$N-Z$ values. The values of $N-Z$ are marked in the figure, 
next to the corresponding chains.    
}
\label{fig1}
\end{figure}

Nuclear structure is extensively studied in relation to
mean-field properties, by analyzing nuclear masses, binding energy,       
shell effects or deformation.
Additional insight on nuclear structure is carried by                    
other frequently investigated observables; among these are
the yields of the residues in low energy fission. In this
case, the fragment distribution reveals an enhanced
production of the even elements, which gradually vanishes                 
with increasing reaction energy.
The disappearence of this staggering with the excitation
energy seemed to constrain the study of nuclear structure to systems      
with low excitation energies.                                             
Though, some experiments dedicated to different and more
violent reactions, like spallation or fragmentation,
revealed similar structures in the yields of the residues~\cite{Ric02}.
A very complete systematics of structural effects in the
isotopic distributions of highly excited systems is the
result of a recent experiment: the residue cross-sections of
the reaction $^{56}$Fe+p and $^{56}$Fe+$^{\textrm{nat}}$Ti at 1~$A$~GeV
were measured in inverse kinematics with the FRagment                     
Separator at GSI (Darmstadt).
In fig.~(\ref{fig1}) the cross sections are presented
ordered according to different chains of isotopes with given
$N-Z$, for even (left) and odd (right) masses, respectively.
Even-mass isotopes manifest an enhanced production of even
elements all along the different chains.
The staggering is maximum for symmetric nuclei ($N=Z$), and it
gradually smooths down for more asymmetric isotopes.
The case of odd masses is more complex: proton-rich isotopes              
(the chain $N-Z=-1$) show an enhanced production of even
elements, while the staggering reverses in favour of an
enhanced production of odd elements for neutron-rich nuclei.
The $^{56}$Fe+$^{\textrm{nat}}$Ti system, introducing appreciably                  
higher excitation energy than the $^{56}$Fe+$p$ system                       
on the average, shows higher
cross sections, but identical features in the staggering along            
the chains of given $N-Z$.
From this comparison, and from the extension to other
measured highly excited systems~\cite{Ric02},
we conclude that the observed structure effect
does not depend on the increase of excitation energy, and it
reveals to be a general property of spallation and
fragmentation residues.
%
%
%

\section{A schematic explanation}
%
%
\begin{figure}[b!]
\begin{center}
\includegraphics[angle=-90, width=1\textwidth]{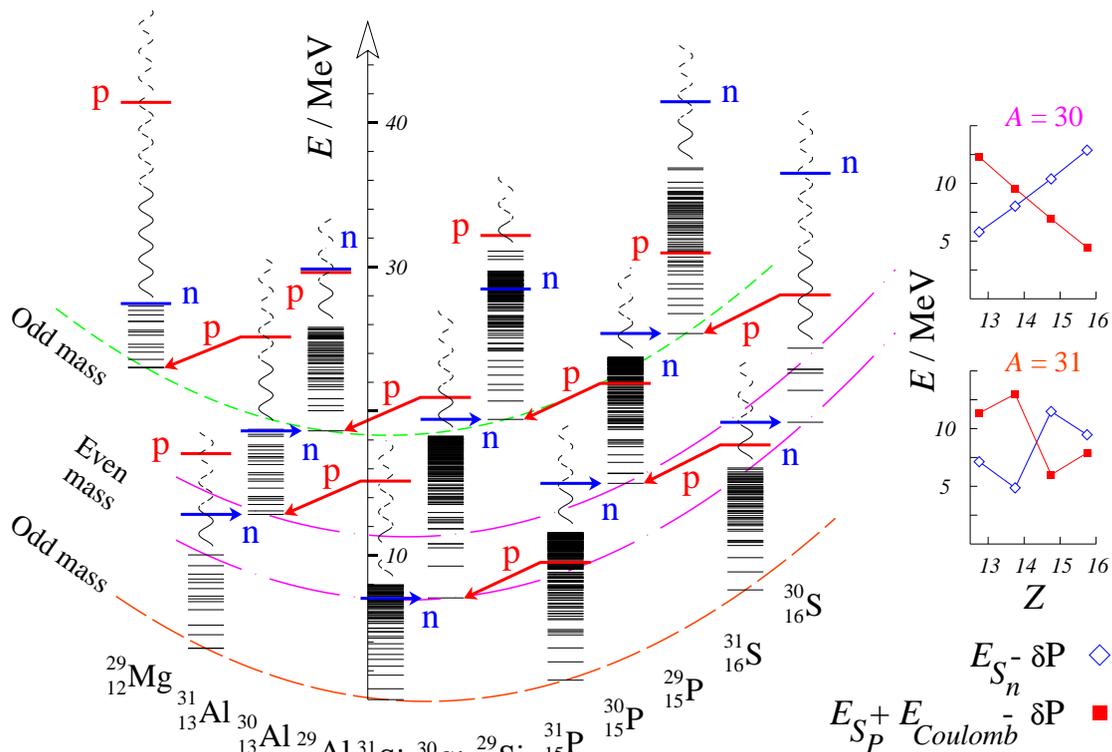}
\end{center}
\caption
{
Evaporation scheme. The experimental levels of a
set of nuclei are ordered on their mass-excess parabolae.
proton and neutron separation energies are marked with "p"
and "n", respectively. On the right, the proton ($S_{p}$) and                 
neutron ($S_{n}$) separation energies,
shifted by the pairing gap $\delta$P, are presented for                       
$A$=30 and $A$=31.
}
\label{fig2}
\end{figure}
A simple statistical evaporation model, where the nuclear
level densities are calculated according to the                              
Fermi-gas model~\cite{Str58} would be sufficient to                          
reproduce all the features observed in the yields, in first
order~\cite{Ric02}.
This could seem to be in contradiction with the                              
counterbalancing of the pairing gap in the nuclear masses
and in the level densities.
On the contrary, in each evaporation step, the probability
of the possible decay channels do not only reflect the level
densities of the daughter nucleus, but they also depend on
the number of excited levels of the mother nucleus that
could decay into the daughter.
The excited levels available for the decay extend from the
separation energy of the daughter nucleus down to the
separation energy of the mother nucleus, increased of the
Coulomb barrier in the case of charged particle emission.
The separation energy of the mother nucleus corresponds to
the ground state of the daughter nucleus.
This is sketched in fig.~(\ref{fig2}), where the levels
of some isotopes are distributed on their mass-excess
parabolae.
Let us consider the case of an odd-mass nucleus decaying
into an even-mass nucleus.
A series of even-mass isotopes ($^{30}$Al, $^{30}$Si,
$^{30}$P, $^{30}$S) show a smooth variation of the
separation energies as a function of the element, once
shifted by the pairing gap $\delta$P.                                        
The absence of staggering in the separation energies is
reflected in a smooth variation of the level density for the
even-mass nuclei as a function of the element.
Nevertheless, due to the pairing gap, odd-mass nuclei
decaying into even-even daughters ($^{30}$Si or $^{30}$S)
have more excited levels available for the decay with respect                
to odd-mass nuclei decaying into odd-odd daughters
($^{30}$Al or $^{30}$P).
At the very end of the evaporation process, the decay in the
ground state of the daughter nucleus becomes so relevant to
determine the overproduction of even-even nuclei compared                    
to odd-odd ones.
A slightly different discussion should be dedicated to the
formation of odd-mass residues ($^{29}$Mg, $^{29}$Al,
$^{29}$Si, $^{29}$P). As the ground states of odd-mass
nuclei are all ordered along the same mass parabola, the
restoring of the structure in the production yields should
be determined by the separation energy that shows up as                      
an even-odd staggering in both, proton and neutron                           
separation energies, however with different signs,                           
depending on the neutron excess.
In odd-mass neutron-rich nuclei ($^{29}$Mg, $^{29}$Al) the
neutron separation energy, that is lower than the proton
separation energy, determines the choice of the most
probable evaporation channel. Thus, the residues will
reflect the structure of the neutron separation energy
favouring the production of odd elements.
Contrarely, the yields of odd-mass proton-rich nuclei
($^{29}$Si, $^{29}$P) reflect the structure of the proton
separation energy favouring the production of even elements.

\section{Conclusions and open questions}
We demonstrated qualitatively that the structure observed in
the nuclide cross section of spallation and fragmentation residues           
is a result of the very last steps of the evaporation
process.
It should be pointed out that the strength of the staggering
is remarkably high. $A$ study based on Tracy's analysis~\cite{Tra72}
would reveal a strenght higher than 50\% for the even-odd
staggering of the $N=Z$ chain, and up to 20\% for the
odd-even staggering of the odd-mass neutron rich nuclei.
This is to be compared to the even-odd staggering that
characterizes the low-energy fission yields, measured to
reach a strength of around 40\% at maximum~\cite{Ste98}.
Another interesting aspect is the much higher production of
alpha-multiple nuclei (i.e. the huge staggering along the
$N=Z$ chain). 
This could be understood as an effect of the lower binding of 
odd-odd symmetric nuclei due to the effect of the Wigner term.
Nevertheless, if we add the remark that the hot fragments
of the reaction $^{56}$Fe+$p$ or $^{56}$Fe+$^{\textrm{nat}}$Ti could
have spent a considerable part of their excitation energy
undergoing a multifragmentation-like or break-up
process~\cite{Nap03}, we might also consider alpha-cluster                         
emission as an additional channel responsible of the
restoring of the structural effects in the production
yields.


\begin{thebibliography}{0}
\bibitem{Ric02}
M. V. Ricciardi et Al,
in {\it Symp. on Nucl. Clusters}
(Rauischholzhausen, Germany, 2002), and referencess therein.
%
\bibitem{Tra72}
B. L. Tracy et Al,
{\it Phys. Rev} {\bf C5}, 222 (1972).
%
\bibitem{Str58}
V. M. Strutinski,
in {\it Int. Conf. on Nuclear Physics},
p. 617 (Paris, Italy, 1958).
%
\bibitem{Ste98}
Steinhaeuser et Al,
{\it Nucl. Phys.} {\bf A634}, 89 (1998).
%
\bibitem{Nap03}
P. Napolitani et Al,
in {\it XLI Int. Winter Meeting on Nucl. Phys.}
(Bormio, Italy, 2003).
%
\end{thebibliography}
\end{document}